\PassOptionsToPackage{normalem}{ulem}
\documentclass[lettersize,journal]{IEEEtran}
\usepackage{amsmath,amsfonts}
\usepackage{mathtools}
\usepackage{amsmath}
\usepackage{xcolor}
\usepackage{array}
\usepackage{soul}
\setstcolor{red}
\usepackage{tabularray}
\usepackage{textcomp}
\usepackage{stfloats}
\usepackage{url}
\usepackage{verbatim}
\usepackage{stfloats}
\usepackage{cite}
\usepackage{xspace}
\usepackage{enumitem}
\usepackage{amsfonts}
\usepackage{nicematrix}
\usepackage{ulem}
\usepackage{tikz}
\usepackage{amssymb}
\usepackage{float}
\usepackage{tabularx}
\usepackage{bm}
\usepackage{placeins}
\usepackage{algorithm}
\usepackage{algorithmic}
\usepackage{subcaption}
\usepackage{graphicx}
\usepackage{xcolor}
\usepackage[normalem]{ulem}
\usepackage[printonlyused]{acronym}
\usepackage[acronym]{glossaries}
\usepackage{xcolor}
\usepackage{mathtools}

\newacro{ISAC}{integrated sensing and communication}
\newacro{3GPP}{3rd Generation Partnership Project}
\newacro{SNR}{signal-to-noise ratio}
\newacro{SINR}{signal-to-interference-noise ratio}
\newacro{MIMO}{multiple-input multiple-output}
\newacro{TRPs}{transmit and receive points}
\newacro{ICT}{information communication technology }
\newacro{GHG}{greenhouse gas}
\newacro{ITU}{International Telecommunication Union}
\newacro{GSMA}{Global System for Mobile Communications Association}
\newacro{GeSI}{Global e-Sustainability Initiative}
\newacro{SBTi}{Science Based Targets initiative }
\newacro{RF}{radio-frequency}
\newacro{ES}{energy saving}
\newacro{BS}{base station}
\newacro{CU}{communication user}
\newacro{AWGN}{additive white Gaussian noise}
\newacro{DFRC}{dual-functional radar communication}
\newacro{GA}{genetic algorithm}
\newacro{MINLP}{mixed-integer
nonlinear program}
\newacro{TDD}{time division duplex}
\newacro{MUSIC}{multiple signal classification}
\newacro{SA}{sparse array}
\newacro{QoS}{Quality of Service}
\newacro{RIS}{reconfigurable intelligent surface}
\newacro{V2I}{vehicle-to-infrastructure}
\newacro{CSI-RS}{channel state information reference signal}
\newacro{RCS}{radar cross section}
\newacro{SDP}{semidefinite program}

\title{Posterior-Confidence Driven Beamforming for Energy-Efficient Integrated Sensing and Communication}
\author{Nusaibah A. Alshorman, and H\"{u}seyin Arslan,~\IEEEmembership{Fellow,~IEEE}
\thanks{N. Alshorman and H. Arslan are with the Department of Electrical and Electronics Engineering, Istanbul Medipol University, Istanbul, Turkey. E-mail: nusaibah.abusanad@std.medipol.edu.tr, huseyinarslan@medipol.edu.tr.}}
\begin{document}

\maketitle

\begin{abstract}
Energy efficiency will pose an essential limitation for sixth-generation (6G) integrated sensing and communication (ISAC) systems, given the high sensing power consumption associated with persistent sensing, despite stable communication requirements. This paper proposes an energy-efficient multiple-input multiple-output (MIMO) dual-functional radar-communication (DFRC) beamforming framework that minimizes transmit power while guaranteeing per-user signal-to-interference-plus-noise ratio (SINR) and reliable multi-target tracking. The key innovation is a tracking-aware, skip-enabled sensing policy that departs from the conventional always-on probing paradigm. Instead of enforcing sensing at every epoch, sensing is selectively triggered according to two complementary statistics derived from an extended Kalman filter (EKF): a posterior confidence metric and the normalized innovation squared (NIS). While the former ensures accurate estimation, the latter guarantees reliable measurements, and thus sensing can only be activated when additional information is required. To ensure robustness under intermittent sensing, sector-based beampattern constraints are combined with a nonzero safety illumination floor imposed to guarantee reliable target tracking when skipping occurs. Numerical results show that the proposed framework achieves a significant reduction in transmit power compared to other baselines, without any deterioration in the communication system's performance or excessive impact on the sensing process.

\end{abstract}

\begin{IEEEkeywords}
Beamforming design, Extended Kalman filter, Energy efficient, ISAC.
\end{IEEEkeywords}

\section{Introduction}
\IEEEPARstart{F}{uture} Information and Communication Technology (ICT) systems contribute to greenhouse-gas emissions, estimated between 1.8\% and 2.8\% \cite{freitag2021real}, and even go up to 3.9\% in other reports \cite{belkhir2018assessing}. Accordingly, international efforts led by the \ac{ITU} \cite{ITU2020}, \ac{GSMA} \cite{GSMA2023}, \ac{GeSI} \cite{GeSI2017ICTGuidance}, and the \ac{SBTi} \cite{ITU2020ScienceTargets} have been conducted to reach net-zero ICT  emissions by 2050. On top of that, \ac{3GPP} continues to standardize practical mechanisms to reduce network energy consumption while preserving the quality of service (QoS) experienced by the served users. These developments highlight that future wireless systems must not only deliver higher performance but also operate under strict energy-efficiency constraints.\par

Integration of sensing and communication (ISAC) within a unified framework is expected to be a main characteristic of the sixth generation (6G). While this integration enables efficient spectrum utilization, this coexistence introduces a fundamental trade-off: aggressive sensing operations significantly increase energy consumption and can degrade communication throughput, whereas insufficient sensing compromises tracking reliability \cite{kaushik2024toward,10663788,valiulahi2023net}. Consequently, energy-efficient (EE) ISAC design must jointly optimize sensing accuracy and communication performance under stringent power budgets.

Recent studies have extended EE-ISAC toward networked and aerial systems, where resources are jointly optimized under energy constraints. For instance, UAV-assisted and distributed ISAC frameworks investigate the trade-off between sensing accuracy and network-level energy consumption \cite{zhu2024resource,zhou2025trade,meng2022uav,khalili2023uavisac}. While these works provide insights into energy-aware scheduling, they primarily focus on distributed architectures rather than base-station-centric tracking. Another line of research targets EE beamforming and transceiver design, including IRS-assisted configurations, NOMA-based joint sensing, and mobility-aware predictive beamforming \cite{yu2024irs,11107235}. Furthermore, EE-ISAC has been studied through various resource allocation frameworks, such as power allocation for Internet-of-Vehicles (IoV) \cite{10506471}, MIMO designs with non-transmission power modeling \cite{wu2023eemimo}, (active-)RIS enhancement \cite{rihan2024activeris}, multi-BS coordination \cite{cui2025twc}, antenna selection \cite{11073166}, and cell-free architectures \cite{behdad2024cellfree}. 

Despite these advances, most existing approaches focus on spatial or architectural optimization while implicitly assuming continuous sensing operation. This assumption leads to unnecessary energy expenditure, particularly in tracking scenarios where target states can be predicted with high confidence. In such applications, EE is critical as the system must maintain communication quality-of-service (QoS) and sensing accuracy without frequent, costly re-acquisition procedures \cite{10445319}, \cite{11129138}. While tracking performance is typically quantified via the Cramér-Rao bound (CRB) \cite{10217169}, preserving fidelity under reduced activity remains a challenge. In particular, the temporal dynamics of sensing necessity are largely overlooked, namely, \emph{when} sensing operations should be activated or skipped based on the instantaneous tracking uncertainty. This limitation is caused by the implicit assumption that sensing is always beneficial to do regardless of the quality of the current estimate. Nevertheless, in practical tracking systems, the estimation error covariance provides a direct measure of confidence, which can be exploited to reduce unnecessary sensing updates. By ignoring this temporal redundancy, existing approaches fail to fully leverage the predictive capability of tracking filters, leading to inefficient energy usage without proportional gains in sensing performance.

From a standard point of view, 3GPP New Radio (NR) offers sufficient flexibility to support an adaptive sensing process through aperiodic \ac{CSI-RS} triggering \cite{3gpp38214}. Conversely, 3GPP Network Energy Saving (NES) mechanisms define sleep states and energy models that enable translating reduced sensing activity into tangible energy savings \cite{3gpp38864}. Furthermore, ISAC-related study items specify key performance metrics such as accuracy, update rate, reliability, and confidence levels \cite{3gppTR22837R19}. These developments highlight that practical ISAC systems must address not only \emph{how to sense} but, more importantly, \emph{when to sense} under energy constraints. However, existing works do not explicitly incorporate such adaptive sensing decisions into the physical-layer beamforming design \cite{10680586, 10445319}, limiting their applicability in practical energy-constrained deployments. This trade-off between sensing frequency and power consumption becomes critical in applications where sensing needs to be adaptive based on the uncertainty associated with the target state. In such cases, continuous sensing may lead to redundant measurements with limited information gain, especially when the target dynamics are smooth and predictable. On the other hand, reducing sensing probing without proper control can result in loss of observability and degraded estimation accuracy. Therefore, an effective ISAC design must account for both the spatial and temporal dimensions of sensing, ensuring that sensing resources are allocated not only where needed but also when needed.\par

Even though Extended Kalman filter (EKF) based tracking has been widely adopted in radar and ISAC systems and has recently been integrated into predictive beamforming and cooperative sensing frameworks, existing approaches primarily focus on performance enhancement, such as beam alignment accuracy and tracking robustness \cite{yang2025cooperative, zhong2025joint, pang2024dynamic}. In addition, these studies do not explicitly address adaptive sensing activation decisions (when and how often the sensing is required), which becomes crucial for controlling energy consumption while satisfying sensing requirements in practical deployments. This gap motivates the need for a unified framework that jointly integrates tracking confidence, sensing activation, and transmit beamforming under energy constraints. 

Inspiring by these observations, this paper proposes a posterior-aware sensing activation framework for energy-efficient ISAC tracking.  We propose an EKF-driven activation policy that enables sensing operations skipping during high-confidence periods. The base station (BS), by specifying a "Sensing Sleep" policy, is capable of dynamically skipping (shutting down probing periods) when the status of the target is predicted with sufficient confidence. Unlike conventional EKF schemes that only determine when to update, the proposed approach incorporates sensing activation decision within a convex transmit covariance design. To this end, we incorporate an EKF posterior confidence and normalized innovation squared (NIS) gating into a beamforming optimization problem that minimizes power consumption while satisfying communication and sensing constraints. Additionally, we implement a "Safety Illumination Floor" for our system, guaranteeing some level of observability even during intervals that can be skipped. This mechanism prevents costly track re-acquisition while maximizing energy savings. At the same time, per-user SINR constraints are enforced to mitigate radar-to-communication interference, ensuring that the "Green" policy does not affect the primary communication. Overall, the proposed framework turns sensing from a continuous into an adaptive, confidence-driven process, enabling substantial energy savings while maintaining reliability of the system, and providing a realistic path toward green ISAC implementations in future 6G networks. In this regards, the main contributions of this paper are summarized as follows:

\begin{itemize}

\item We introduce an energy-aware ISAC beamforming framework, which takes into account the temporal sensing activation while performing the transmit covariance matrix design. 

\item We develop a posterior-aware sensing activation mechanism based on EKF, where sensing is governed by estimation confidence as well as the normalized innovation squared (NIS) gating.

\item We propose a joint transmit covariance optimization problem with the objective of minimizing transmit power while ensuring per-user SINR constraints and target illumination requirements. 

\item To ensure robust tracking under intermittent sensing, we introduce a safety illumination constraint that guarantees minimum target observability, preventing track divergence while maintaining energy savings.

\end{itemize}

The remainder of this paper is organized as follows. Section II presents the ISAC system architecture, including the channel and signal models for both communication and sensing. Section III introduces the tracking model and the adopted performance metrics. Section IV formulates the optimization problem, while Section V describes the proposed skip-aware beamforming framework. Section VI provides numerical results and discussion, and Section VII concludes the paper.

Throughout the paper, capital boldface letters denote matrices, and lowercase boldface letters indicate vectors. The notation $\mathbf{A} \succeq \mathbf{0}$ means that $\mathbf{A}$ is a positive semi-definite matrix. The Hermitian, trace, and transpose operators are denoted by $(.)^H$, $\text{tr}(.)$, and $(.)^T$  respectively. 

\section{System Model}

\begin{figure}[!t]
    \centering
   \includegraphics[width=3in]{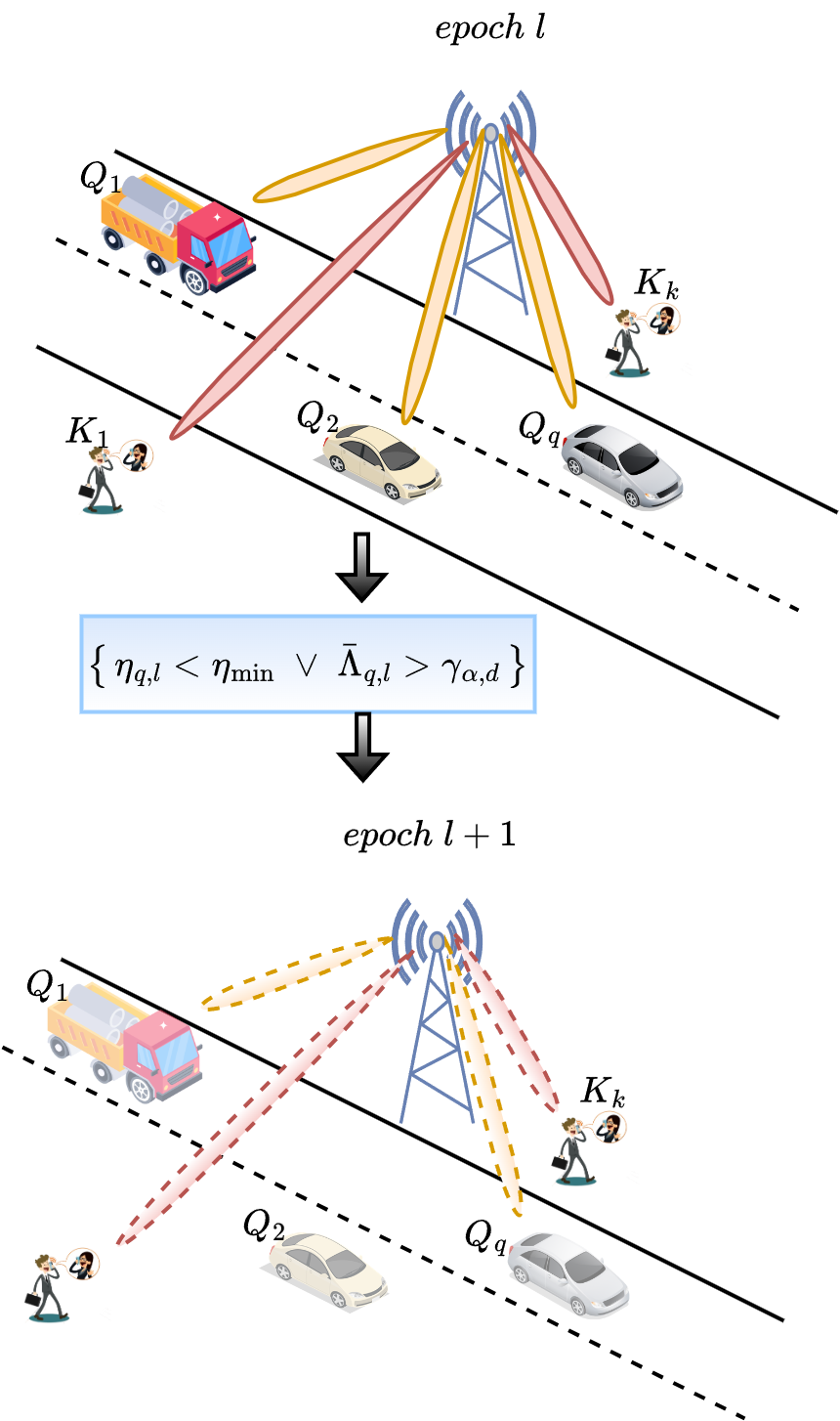}
    \caption{System model.}
    \label{System model}
\end{figure}

We consider a MIMO \ac{DFRC} \ac{BS} equipped with $N_t$ transmit antennas and $N_r$ receive antennas, which serves $K$ downlink single-antenna communication users while tracking $Q$ point targets. The set of communication users is indexed by $\mathcal{K}=\{1,2,\ldots, K\}$, while the set of targets is indexed by $\mathcal{Q}=\{1,2,\ldots, Q\}$ as depicted in Fig. \ref{System model}. During the ISAC transmission period, the BS precodes the ISAC signal. The total transmission time $T_s$ is divided into $l$ epochs. At the beginning of each epoch, the wireless channel state information is updated and assumed to remain unchanged throughout the epoch. Each epoch corresponds to a slow-time decision interval in which both beamforming and sensing activation are jointly optimized. Within an epoch, the designed beamformers are reused across multiple transmission slots, which reflects practical NR-based operation where control decisions are updated at a slower timescale than symbol-level transmission. To achieve satisfactory communication and sensing performance, we exploit the maximum DoFs provided by MIMO and transmit $M$ additional probing streams.The transmit signal at each slot is \cite{liu2021cramer, 10556732, 9724205}
\begin{equation}
  \mathbf{X}_l = \mathbf{W}_{c,l}\,\mathbf{S}_{c,l} + \mathbf{W}_{r,l}\,\mathbf{S}_{r,l},
  \label{eq:X}
\end{equation}
where $\mathbf{X}_l\in\mathbb{C}^{N_t\times L}$ is the transmit signal matrix with frame length $L>N_t$.
The $i$th communication data stream is denoted by $\mathbf{s}_{i,l}$ $\forall$   $i\in\mathcal{K}$, and $\mathbf{S}_{c,l} = [\mathbf{s}_{1,l}^{\text{T}}, \dots, \mathbf{s}_{K,l}^{\text{T}}]^{\text{T}} \in \mathbb{C}^{K\times L}$ contains the $K$ date streams intended for the $K$ users. Similarly, the $m$th radar probing stream is denoted by $\mathbf{s}_{C+m,l}\in\mathbb{C}^{1\times L}$ for $m\in\{1,2,\ldots,M\}$, and
$ \mathbf{S}_{r,l}=  
[\mathbf{s}_{C+1,l}^{\text{T}},
  \dots, \mathbf{s}_{C+M,l}^{\text{T}}]
  \in\mathbb{C}^{M\times L}$ contains $M$ individual probing  streams with $M < N_t- K$. The matrices $\mathbf{W}_{c,l} = [\mathbf{w}_{1,l},\ldots,\mathbf{w}_{K,l}] \in \mathbb{C}^{N_{t}\times K}$ and $\mathbf{W}_{r,l} = [\mathbf{w}_{K+1,l},\ldots,\mathbf{w}_{K+M,l}] \in \mathbb{C}^{N_{t}\times M}$ contain the transmit beamforming vectors for the data streams and the probing streams, respectively.

We assume that both the communication signals and radar probing signals are wide-sense stationary stochastic processes with zero-mean and unit power. The communication data signals for different users are uncorrelated, so $\frac{1}{L}\mathbb{E}[\mathbf{S}_{c,l}\mathbf{S}_{c,l}^{H}] = \mathbf{I}_{C}$. The radar probing signals are pseudo-random sequences with zero-mean and unit variance, and are uncorrelated with each other, i.e., $\frac{1}{L}\mathbb{E}[\mathbf{S}_{r,l}\mathbf{S}_{r,l}^{H}] = \mathbf{I}_{M}$. The communication signals and radar probing signals are assumed to be uncorrelated, namely $\mathbb{E}[\mathbf{S}_{c,l}\mathbf{S}_{r,l}^{H}] = \mathbf{0}$.

\subsection{Multiuser Communication Model}
For downlink communications, the signal received at the $i$th
user at epoch $l$ is expressed as
\begin{equation}
\small
\mathbf{y}_{i,l} = \mathbf{h}_{i,l}^{H} \mathbf{w}_{i,l} s_{i,l} + \sum_{j=1,\, j\neq i}^{K}
\mathbf{h}_{i,l}^{H} \mathbf{w}_{j,l} s_{j,l}+
\sum_{j=K+1}^{K+M} \mathbf{h}_{i,l}^{H} \mathbf{w}_{j,l} s_{j,l}+ \mathbf{z}_{i,l},
\label{1}
\end{equation}
where $\mathbf{h}_{i,l} \in \mathbb{C}^{N_t\times 1}$ is the channel
 between the BS and the $i$th user,
modeled as 
$\mathbf{h}_{i,l}=\beta_{i,l}\mathbf{a}(\theta_i)$, where $\mathbf{a}(\theta_i)$ is the ULA steering
vector with $[\mathbf{a}(\theta_i)]_n=\exp(-j\frac{2\pi}{\lambda}nd\cos\theta_i)$, $n=0,\ldots,N_t-1$.
The complex gain $\beta_{i,l}$ captures large and small-scale effects,
and $\mathbf{z}_{i,l} \sim \mathcal{CN} (0, \sigma_i^{2} \mathbf{I}_L)$ is an additive
white Gaussian noise (AWGN) with zero-mean and
covariance $\sigma_i^{2} \mathbf{I}_L$. Thus, the \ac{SINR} of the $i$th user is 
\begin{equation}
\gamma_{i,l} =\;
\frac{\left|\mathbf{h}_{i,l}^{H}\mathbf{w}_{i,l}\right|^{2}}
{\displaystyle \sum_{\substack{j=1,  j\neq i}}^{K} \left|\mathbf{h}_{i,l}^{H}\mathbf{w}_{j,l}\right|^{2}
+ \sum_{j=K+1}^{K+M} \left|\mathbf{h}_{i,l}^{H}\mathbf{w}_{j,l}\right|^{2}
+ \sigma_{i,l}^{2}}.
\tag{5}
\end{equation}

We quantify the multi-user MIMO downlink performance by the average sum rate, defined as
\begin{equation}
r_{K,l} = \frac{1}{K}\sum_{i=1}^{K} \log\!\left(1+\gamma_{i,l}\right).
\end{equation}

\subsubsection{Energy Efficiency}

 We explicitly evaluate the energy efficiency (EE) of the ISAC system, defined as the number of successfully delivered information bits per unit transmit energy. For epoch $l$, the EE is defined as
\begin{equation}
\mathrm{EE}_l = \frac{r_{K,l}}{P_l},
\end{equation}
where $P_l = \mathrm{tr}(\mathbf{R}_{X,l})$ denotes the total transmit power at epoch $l$, with $\mathbf{R}_{X,l} = \mathbf{W}_{c,l}\mathbf{W}_{c,l}^H + \mathbf{W}_{r,l}\mathbf{W}_{r,l}^H$ being the transmit covariance matrix. Therefore, the average energy efficiency over the observation horizon is then given by
\begin{equation}
\mathrm{EE} = \frac{1}{L_e}\sum_{l=1}^{L_e} \mathrm{EE}_l,
\end{equation}
where $L_e$ is the total number of epochs.

\subsection{Radar Sensing Model}
For the radar, the reflected echo signal received at the BS can be expressed in matrix form as 
\begin{equation}
    \textbf{Y}_{R,l} = \textbf{G}_l \textbf{X}_l+\textbf{Z}_{R,l},
\end{equation}
where \( \mathbf{Z}_{R,l} \in \mathbb{C}^{N_r \times L} \) is the AWGN
with the variance of each entry being \(\sigma^2_R \),
and the target response matrix \( \mathbf{G}_l \)
is expressed as the superposition of the response of the individual 

\begin{equation}
  \mathbf{G}_l = \sum_{q=1}^{Q} \beta_q\,\mathbf{b}_l(\theta_q)\,\mathbf{a}_l^H(\theta_q),
\end{equation}
where $\mathbf{a}_l(\theta_q)\in\mathbb{C}^{N_t\times 1}$ and $\mathbf{b}_l(\theta_q)\in\mathbb{C}^{N_r\times 1}$ denote the transmit and receive steering vectors. To decouple spatial beamforming from time–frequency allocation, the time and bandwidth resources are assumed fixed over the tracking interval, such that the complex scattering coefficient $\beta_q$ captures all non-spatial effects (including path loss, \ac{RCS}, range delay, and Doppler) and is approximately constant within each observation epoch.

\section{Tracking Model and Performance Metrics}
To track the motion states of the $Q$ targets at the epoch rate, we adopt an Extended Kalman Filter (EKF).
Throughout an epoch, we assume that any pair of targets can be resolved in at least one domain range, angle, or Doppler, so that a reliable data association is feasible.
\subsection{EKF Tracking Model}
The EKF maintains a Cartesian constant-velocity state for target $q$ at epoch $l$,
\begin{equation}
    \boldsymbol{\chi}_{q,l} =
    \begin{bmatrix}
        x_{q,l} & y_{q,l} & v_{x,q,l} & v_{y,q,l}
    \end{bmatrix}^{\mathrm{T}}.
\end{equation}
Under a constant-velocity model with sampling interval $\Delta T_l$,
\begin{equation}
    \boldsymbol{\chi}_{q,l} = \mathbf{D}(\Delta T_l)\,\boldsymbol{\chi}_{q,l-1} + \mathbf{w}_{q,l},
\end{equation}
where $
\mathbf{w}_{q,l}\sim\mathcal{N}(\mathbf{0},\mathbf{Q}_w(\Delta T_l))$, $\mathbf{D}$ modeled as 

\begin{equation}
\mathbf{D}=
\begin{bmatrix}
1 & 0 & \Delta T_l & 0 \\
0 & 1 & 0 & \Delta T_l \\
0 & 0 & 1 & 0 \\
0 & 0 & 0 & 1
\end{bmatrix},
\end{equation}
And $\mathbf{Q}_w$ represented as 
\begin{equation}
\mathbf{Q}_w= q_a
\begin{bmatrix}
\frac{\Delta T_l^3}{3} & 0 & \frac{\Delta T_l^2}{2} & 0 \\
0 & \frac{\Delta T_l^3}{3} & 0 & \frac{\Delta T_l^2}{2} \\
\frac{\Delta T_l^2}{2} & 0 & \Delta T_l & 0 \\
0 & \frac{\Delta T_l^2}{2} & 0 & \Delta T_l
\end{bmatrix}.
\end{equation}

Given the probing $\mathbf{X}_l$ and sensing response $\mathbf{G}_l$, the radar forms nonlinear measurements of angle, range, and Doppler. Denoting the measurement of target $q$ at epoch $l$ by
\begin{equation}
    \mathbf{z}^{\mathrm{rad}}_{q,l} = g(\boldsymbol{\chi}_{q,l}) + \mathbf{v}_{q,l},\qquad
    \mathbf{v}_{q,l}\sim\mathcal{N}(\mathbf{0},\mathbf{R}_z),
\end{equation}
where $g(\cdot)$ maps state to observables,
$\mathbf{J}_{q,l}= \left.\frac{\partial g}{\partial \boldsymbol{\chi}}\right|_{\hat{\boldsymbol{\chi}}_{q,l|l-1}}$.
 The prediction and correction at each epoch $l$ are
\begin{align}
\text{Prediction:}&\quad
\hat{\boldsymbol{\chi}}_{q,l|l-1} = \mathbf{D}\hat{\boldsymbol{\chi}}_{q,l-1|l-1}, \label{Pr1}\\
&\quad\mathbf{P}_{q,l|l-1} = \mathbf{D}\mathbf{P}_{q,l-1|l-1}\mathbf{D}^{\mathrm{T}} + \mathbf{Q}_w,\label{Pr2}\\
\text{Update:}&\quad
\mathbf{S}_{q,l} = \mathbf{J}_{q,l}\mathbf{P}_{q,l|l-1}\mathbf{J}_{q,l}^{\mathrm{T}} + \mathbf{R}_z,\\
&\quad \mathbf{L}_{q,l} = \mathbf{P}_{q,l|l-1}\mathbf{J}_{q,l}^{\mathrm{T}}\mathbf{S}_{q,l}^{-1},\\
&\quad \hat{\boldsymbol{\chi}}_{q,l|l} = \hat{\boldsymbol{\chi}}_{q,l|l-1}
+ \mathbf{L}_{q,l}\big(\mathbf{z}^{\mathrm{rad}}_{q,l} - g(\hat{\boldsymbol{\chi}}_{q,l|l-1})\big),\\
&\quad \mathbf{P}_{q,l|l} = (\mathbf{I}_4 - \mathbf{L}_{q,l}\mathbf{J}_{q,l})\mathbf{P}_{q,l|l-1}.
\end{align}
\subsection{Performance Metrics}
\subsubsection{Confidence and NIS Metrics}
To avoid misassociated updates, we gate measurements via the Normalized Innovation Squared (NIS)
\begin{equation}
\Lambda_{q,l}= \tilde{\mathbf{y}}_{q,l}^{\top}\mathbf{S}_{q,l}^{-1}\tilde{\mathbf{y}}_{q,l}.
\label{NIS}
\end{equation}
where $\Lambda_{q,l}$ follows a $\chi^2_d$ distribution with $d$ equal to the number of measurements.
The update is accepted if $\Lambda_{q,l}\le \gamma_{\alpha,d}$, where $\gamma_{\alpha,d}$ is the $\alpha$-quantile of $\chi^2_d$; otherwise a missed-detection (prediction-only) step is performed.
 we define the scalar confidence metric as
\begin{equation}
\eta_{q,l} = \frac{1}{\operatorname{tr}\!\big(\mathbf{P}_{q,l}\big)}.
\label{conf}
\end{equation}
\subsubsection{Tracking Stability Metric}
Let $\mathbf{s}_{q,l} = [\theta_{q,l}, r_{q,l}, v_{q,l}]^{\top}$ denote the true angle, range, and velocity of target $q$ at epoch $l$, and $\hat{\mathbf{s}}_{q,l}$ the corresponding EKF estimate. To quantify tracking performance under intermittent sensing, we define a normalized tracking error indicator as
\begin{equation}
\delta_{q,l} =
\begin{cases}
1, &
\displaystyle
\frac{|\hat{\theta}_{q,l}-\theta_{q,l}|}{\theta_{\max}}
+
\frac{|\hat{r}_{q,l}-r_{q,l}|}{r_{\max}}
+
\frac{|\hat{v}_{q,l}-v_{q,l}|}{v_{\max}}
> 1,\\[6pt]
0, & \text{otherwise},
\end{cases}
\end{equation}
where $\theta_{\max}$, $r_{\max}$, and $v_{\max}$ denote application-specific tolerances that normalize the contributions of different state components and reflect acceptable sensing accuracy. This formulation is consistent with threshold-based track loss criteria commonly adopted in radar tracking systems \cite{song2022trackingmetrics}.

A track for target $q$ is declared lost if the error condition persists for at least $T_{\mathrm{loss}}$ consecutive epochs:
\begin{equation}
\mathcal{L}_q(\rho) =
\begin{cases}
1, & \exists\,l \;\text{such that}\;
\displaystyle
\sum_{t=l}^{l+T_{\mathrm{loss}}-1} \delta_{q,t}
= T_{\mathrm{loss}}, \\[6pt]
0, & \text{otherwise}.
\end{cases}
\end{equation}

The multi-target track-loss probability is defined as
\begin{equation}
P_{\mathrm{loss}}(\rho)
=
\frac{1}{Q}
\sum_{q=1}^{Q}
\mathbb{E}\!\left[\mathcal{L}_q(\rho)\right],
\end{equation}
where $\rho \in [0,1]$ is the safety illumination factor , which defines the minimum fraction of sensing power preserved during skipped epochs. In particular, inactive targets are illuminated with $\rho \alpha_q$, ensuring continued state observability and stable EKF operation. The parameter $\rho$ therefore controls the fundamental trade-off between energy efficiency and tracking stability, the role of $\rho$ in the beamforming design and its impact on sensing activation will be further elaborated in Section~V . This metric provides a system-level measure of tracking stability by jointly capturing angle, range, and velocity degradation over time.\par

\textit{Remark (Communication–Sensing Coupling):}

The transmitted signal $\mathbf{X}_l$ simultaneously carries communication data and sensing information, leading to inherent coupling between the two functionalities. In particular, the radar probing streams introduce additional interference to communication users, while communication beams contribute to the sensing beampattern. This dual effect is explicitly captured in the SINR expression and the sensing covariance design. From an energy-efficiency perspective, this coupling implies that reducing sensing activity (through adaptive probing) directly decreases the total transmit power, while maintaining the required SINR levels ensures that communication performance is not compromised. Therefore, the proposed design leverages this coupling to optimize sensing reliability and communication efficiency jointly. The detailed problem formulation and the proposed algorithm are presented in the subsequent sections.

\section{Problem Formulation}
We first consider a baseline continuous sensing design, where sensing illumination is enforced for all targets at every epoch. 
This formulation serves as a reference design that reflects conventional ISAC systems, where sensing is treated as a mandatory and continuous operation. While such an approach ensures high tracking reliability, it may lead to unnecessary energy expenditure, particularly in scenarios where the target state can be accurately predicted over multiple epochs. Therefore, this baseline will later be used to highlight the energy savings enabled by the proposed skip-aware framework.
 Under this assumption, the downlink ISAC transmit covariances at epoch $l$ are designed to minimize the total transmit power while guaranteeing (i) per-user SINR requirements for communication and (ii) minimum sensing illumination toward $Q$ targets over predefined angular sectors.

The transmit covariance matrix is given by
\begin{equation}
\mathbf{R}_{X,l}=\mathbf{F}_{l}+\sum_{i\in\mathcal{K}}\mathbf{W}_{i,l},
\end{equation}
where $\mathbf{W}_{i,l}=\mathbf{w}_{i,l}\mathbf{w}_{i,l}^{H}$ is the covariance matrix of the $i$th communication beam and
$\mathbf{F}_{l}=\mathbf{W}_{r,l}\mathbf{W}_{r,l}^{H}$ is the sensing covariance matrix. For user $i$, define $\mathbf{H}_{i,l}= \mathbf{h}_{i,l}\mathbf{h}_{i,l}^{H}$.
For target $q$ with nominal direction $\theta_q$, we sample a sector
$\mathcal{M}_q=\{\theta_{q,m}\}$ and define
\begin{equation}
\mathbf{A}_{q,m,l}=\mathbf{a}_l(\theta_{q,m})\,\mathbf{a}_l^{H}(\theta_{q,m}). 
\end{equation}

This sector-based formulation ensures robustness against angular uncertainty and target motion within each epoch, by enforcing a minimum beampattern gain across a set of discrete angular samples rather than a single direction. As a result, the sensing constraint guarantees reliable target illumination even under model mismatch or estimation errors. Therefore, the beamforming design is formulated as
\begin{subequations}\label{eq:P1}
\small
\begin{align}
\min_{\substack{ \{\mathbf{W}_{i,l}\succeq \mathbf{0}\}_{i\in\mathcal{K}}\\ \mathbf{F}_{l}\succeq \mathbf{0} }} \quad
& \sum_{i\in\mathcal{K}} \mathrm{tr}(\mathbf{W}_{i,l}) + \mathrm{tr}(\mathbf{F}_{l})
\label{eq:P1_obj}\\[-2pt]
\text{s.t.}\quad
& \mathrm{tr}(\mathbf{H}_{i,l}\mathbf{W}_{i,l}) \;-\nonumber\\[-2pt]
& \tau_k\big(
      \sum_{\substack{j=1,  j\neq i}}^{K}
      \mathrm{tr}(\mathbf{H}_{i,l}\mathbf{W}_{j,l})
      + \mathrm{tr}(\mathbf{H}_{i,l}\mathbf{F}_{l})
    \big)
\ge \tau_k\,\sigma_i^{2},
\label{eq:P1_sinr}\\[-2pt]
& \mathrm{tr}\!\big(\mathbf{A}_{q,m,l}\big(\mathbf{F}_{l} + \!\sum_{i\in\mathcal{K}} \mathbf{W}_{i,l}\big)\big) \ge \alpha_{q},\quad \forall q\in\mathcal{Q}.
\label{eq:P1_sens}
\end{align}
\end{subequations}

The objective in \eqref{eq:P1_obj} minimizes the total radiated power per epoch. 
Unlike conventional rate-maximization formulations, this problem enforces communication quality-of-service (QoS) through SINR constraints while minimizing transmit power. Consequently, improvements in channel conditions are translated into reduced transmit power rather than increased data rates. The SINR constraints \eqref{eq:P1_sinr} are affine in the covariance variables, and 
the sensing constraint \eqref{eq:P1_sens} enforces a minimum beampattern gain $\alpha_q$ toward each target sector at every epoch. This guarantees continuous target illumination, which is essential for stable EKF updates but may lead to redundant sensing when the tracking uncertainty is low. This observation motivates the development of adaptive sensing strategies, as discussed in the subsequent section. The resulting formulation is a convex semidefinite program (SDP) that can be efficiently solved using standard optimization solvers.

\textit{Remark (Rank-One Recovery):}
Although \eqref{eq:P1} is solved via semidefinite relaxation, the relaxation is tight for the considered structure. This property follows from the covariance-based formulation, where both the objective function and the constraints depend only on the aggregate transmit covariance $\mathbf{R}_{X,l}$ rather than the individual beamforming ranks. Furthermore, since the user channel matrices $\mathbf{H}_{i,l} = \mathbf{h}_{i,l}\mathbf{h}_{i,l}^H$ are rank-one, the problem admits an exact rank-one reconstruction. Therefore, optimal rank-one beamformers can be recovered without performance loss using the constructive method in~\cite[Theorem 4]{liu2021cramer}. 
This guarantees that the obtained solution does not suffer from the typical suboptimality associated with semidefinite relaxation in general beamforming problems.

\section{The Proposed Skip-Aware Beamforming Framework}

To overcome the inefficiency of continuous sensing, we propose a skip-aware beamforming framework that dynamically determines \emph{when} and \emph{how} to perform sensing based on the tracking state. The key idea is to avoid unnecessary probing when the target state is already well estimated, while preserving sufficient illumination to maintain track observability.
We take the sensing decision at the epoch level by redesigning the ISAC beamformers based on two complementary statistics derived from the EKF:
(i) a posterior-confidence scalar $\eta_{q,l}$ and
(ii) a $\chi^2$-gated NIS $\Lambda_{q,l}$.

The posterior-confidence metric $\eta_{q,l}$ reflects the long-term reliability of the state estimate, while the NIS $\Lambda_{q,l}$ captures the instantaneous consistency between predicted and observed measurements. The joint use of these two metrics enables robust sensing decisions by accounting for both accumulated estimation uncertainty and abrupt model mismatches.\par
This enables skipping unnecessary probing for well-tracked targets and 
reallocating power to communications or targets that need tracking,
thereby reducing the total transmit energy. The sensing binary decision for target $q$ in epoch $l+1$ define as

\begin{equation}
u_{q,l+1}
=
\big\{\, \eta_{q,l} < \eta_{\min} \;\lor\; {\Lambda}_{q,l} > \gamma_{\alpha,d} \,\big\}.
\label{binary}
\end{equation}

Therefore, the sensing activates only when the tracking confidence falls below a predefined threshold or when the measurement innovation indicates inconsistency. Otherwise, sensing is temporarily skipped, indicating that the current estimate is sufficiently reliable to sustain prediction-based tracking.
To guarantee a nonzero safety illumination toward a temporarily unserved target and thereby preserve tracking accuracy, the transmit design is posed as the following SDP

\begin{subequations}
\label{eq:PF}
\small
\begin{align}
\min_{ \{\mathbf{W}_{i,l}\succeq \mathbf{0}\}, \mathbf{F}_{l}\succeq \mathbf{0} }
& \sum_{i\in\mathcal{K}} \mathrm{tr}(\mathbf{W}_{i,l}) + \mathrm{tr}(\mathbf{F}_{l}) \label{eq:PF_obj1}\\[4pt]
\text{s.t.}\quad
& \tau_k \Big( \sum_{j\neq i} \mathrm{tr}(\mathbf{H}_{i,l}\mathbf{W}_{j,l}) + \mathrm{tr}(\mathbf{H}_{i,l}\mathbf{F}_{l}) \Big) \ge \tau_k \sigma_i^{2}, \label{eq:P1_sinr}\\[4pt]
& \mathrm{tr}\bigg(\mathbf{A}_{q,m,l}\Big(\mathbf{F}_{l} + \sum_{j\in\mathcal{K}} \mathbf{W}_{j,l}\Big)\bigg) \nonumber \\ 
& \hspace{35pt} \ge \big(u_{q,l} + \rho(1-u_{q,l})\big)\alpha_q, \quad \forall q, \label{eq:skip_sens1}
\end{align}
\end{subequations}

where the sensing constraint \eqref{eq:skip_sens1} captures the key idea of skip-awareness. If $u_{q,l}=1$, full sensing illumination $\alpha_q$ is required to provide correct updates of measurement. Otherwise, if  $u_{q,l}=0$, then a lower but nonzero sensing illumination $\rho \alpha_q$ is kept, guaranteeing the visibility of the target and preventing the EKF from diverging when skipping a measurement epoch. 

While $\rho$ controls the trade-off between power consumption and tracking robustness. Smaller values of $\rho$ yield higher energy savings but may degrade tracking stability, while larger values improve robustness at the cost of increased transmit power, because lower values correspond to a smaller floor illumination power. The overall procedure is summarized in Algorithm~\ref{main}.


\begin{algorithm}[!t]
\caption{ Skip-aware Confidence/NIS-Driven Sensing \& Energy-Minimizing Beamforming}
\label{alg:conf-nis-energy}
\small
\begin{algorithmic}[1]
\STATE \textbf{Inputs at epoch $l$:} $ \{\mathbf{h}_{i,l}\}_{i=1}^K$, $\mathbf{X}_l$, $\mathbf{G}_l$, thresholds $\eta_{\min}$, $\rho$, $\tau_k$, $\gamma_{\alpha,d}$, and $\alpha_q$.
\STATE \textbf{EKF prediction:} for each target $q$, solve \eqref{Pr1} and \eqref{Pr2}.
\STATE \textbf{NIS gating \& update:} for each $q$ with a measurement $\mathbf{z}^{\mathrm{rad}}_{q,l}$, find ${\Lambda}_{q,l}$ using \eqref{NIS}.
\STATE \textbf{Confidence metric:} for each $q$, find $\eta_{q,l}$ using \eqref{conf}
\STATE \textbf{skip decision for epoch $l{+}1$:} find $u_{q,l+1}$ using \eqref{binary}
$\mathcal{A}_{l+1}\!\leftarrow\!\{q:\,u_{q,l+1}=1\}$.
\STATE \textbf{Energy-minimizing redesign:} set sensing power to zero for $q\notin\mathcal{A}_{l+1}$ and solve Problem \eqref{eq:PF}.
\STATE \textbf{Transmit in $l{+}1$:} apply $\mathbf{W}_{c,l+1}, \mathbf{F}_{l+1}$; 
\STATE \textbf{Loop:} $l\!\leftarrow\!l{+}1$ and repeat.
\end{algorithmic}
\label{main}
\end{algorithm}
\subsection{Computational Complexity Discussion}
The proposed SDP in \eqref{eq:P1} is solved at the epoch level, corresponding to a slow-time beam management and tracking update scale, rather than at the symbol or slot level. 
For the considered tracking dynamics, the resulting beamformers are reused over multiple transmission slots within each epoch, making the approach compatible with practical ISAC operation.

From a computational perspective, \eqref{eq:P1} is a semidefinite program with $(K+1)$ matrix variables of dimension $N_t \times N_t$. When solved using interior-point methods, the worst-case complexity scales as
\begin{equation}
\mathcal{O}(K^3 N_t^4 + K N_t^6),
\end{equation}
which is consistent with standard SDP-based beamforming formulations in ISAC systems. While this complexity may be high for large-scale arrays, the epoch-level execution significantly reduces the effective computational burden in practice, as the optimization is performed infrequently relative to the transmission timescale.

\section{Results and Discussion}

In this section, we present numerical results to evaluate the performance of the proposed algorithm.
We consider a DFRC network with a single BS operating at a carrier frequency of \(f_c = 28~\mathrm{GHz}\) and the transmit power \(P_t = 43\) dBm . The transceiver employs a monostatic uniform linear array with \(N_t = N_r = 16\) antennas. Time is partitioned into epochs, each consisting of two NR slots. The scenario includes \(Q = 4\) radar targets, each moving at a constant speed \(v \in [5,10]~\mathrm{m/s}\), and \(K = 5\) downlink communication users. The NIS gate is set to the \(95\%\) confidence quantile \(\gamma_{\alpha,d}\) (i.e., \(\alpha=0.95\)). The SINR threshold for communication users is set to \(\tau_k = 10~\mathrm{dB}\). 
To preserve continuous target illumination, a nonzero safety floor is enforced, 
where the scaling factor is chosen as \(\rho = 0.3\) and the illumination threshold 
for each target is set to \(\alpha_q = 15~\mathrm{dB}\). Finally, the posterior-confidence threshold $\eta_{\text{min}} = 0.85$. We compare the proposed Adaptive Skip-aware Probing (ASP) scheme
against two baselines: (i) a Full-Probing (FP) strategy that solves the convex min-power
optimization without skip-aware activation, and (ii) a Periodic Low-Power Probing (PLP)
baseline, in which sensing is enforced once every $l$ epochs in a periodic manner, thereby isolating the effect of reduced probing frequency without EKF/NIS-driven adaptivity.\par
\subsection{Sensitivity to Activation Thresholds}
\subsubsection{The normalized innovation squared threshold $\gamma_{\alpha,d}$}
The NIS threshold $\gamma_{\alpha,d}$, determined by the confidence level $\alpha$, controls the EKF measurement gating mechanism and directly affects the frequency of accepted sensing updates. Larger $\alpha$ values relax the gating condition, enabling more frequent skipping and lower sensing power, while smaller values enforce stricter consistency checks.

\begin{figure}[!h]
    \centering
\includegraphics[width=.85\linewidth]{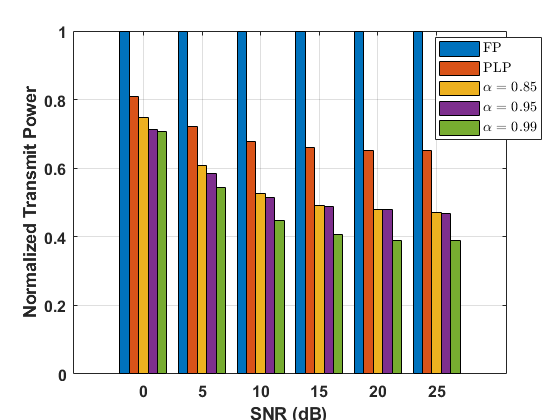}
    \caption{ Normalized transmit power vs SNR for all schemes, $\rho =0.3$ and $\eta_{min} = 0.85$.}
    \label{alpha_pt}
\end{figure}

\begin{figure*}[!t] 
    \centering
    \begin{subfigure}[b]{0.32\textwidth}
        \centering
        \includegraphics[width=\linewidth]{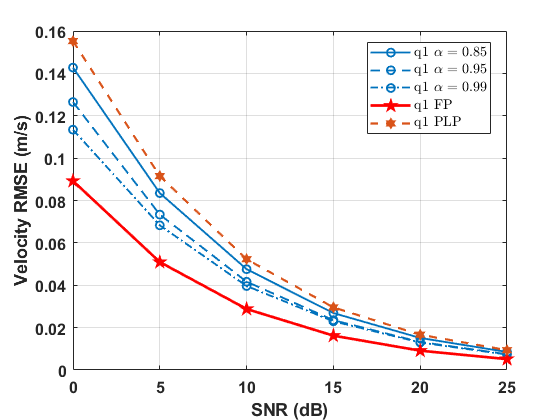}
        \caption{Velocity RMSE.}
        \label{q1_v}
    \end{subfigure}
    \hfill
    \begin{subfigure}[b]{0.32\textwidth}
        \centering
        \includegraphics[width=\linewidth]{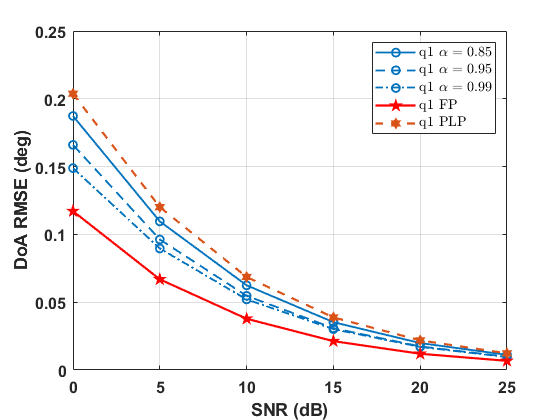}
        \caption{DoA RMSE.}
        \label{q1_doa}
    \end{subfigure}
    \hfill
    \begin{subfigure}[b]{0.32\textwidth}
        \centering
        \includegraphics[width=\linewidth]{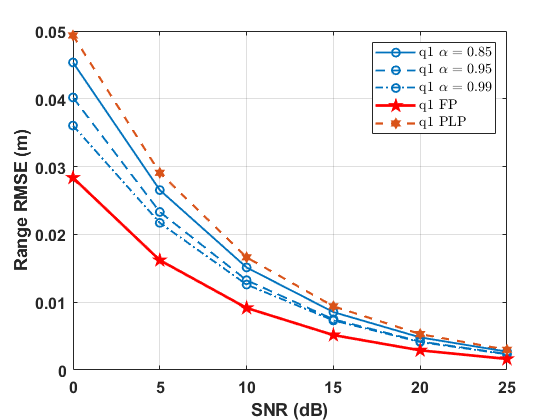}
        \caption{Range RMSE.}
        \label{q1_range}
    \end{subfigure}

    \caption{Q1 performance analysis: RMSE vs SNR for all proposed schemes, $\rho =0.3$ and $\eta_{min} = 0.85$.}
    \label{fig:q1_results}
\end{figure*}

\begin{figure*}[!t] 
    \centering
    \begin{subfigure}[b]{0.32\textwidth}
        \centering
        \includegraphics[width=\linewidth]{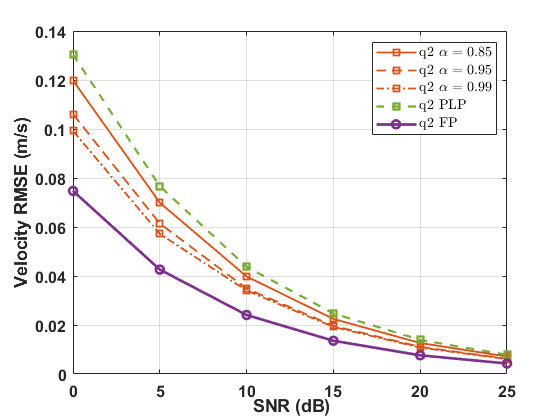}
        \caption{Velocity RMSE.}
        \label{q2_v}
    \end{subfigure}
    \hfill
    \begin{subfigure}[b]{0.32\textwidth}
        \centering
        \includegraphics[width=\linewidth]{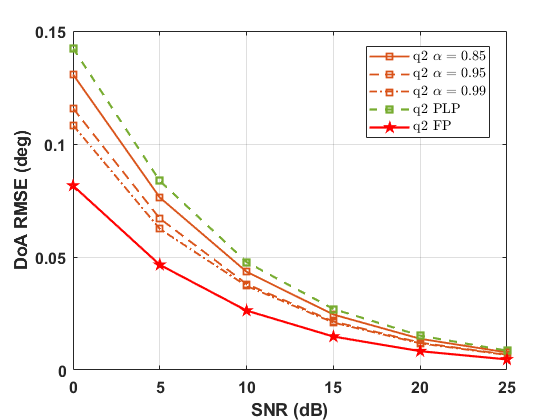}
        \caption{DoA RMSE.}
        \label{q2_doa}
    \end{subfigure}
    \hfill
    \begin{subfigure}[b]{0.32\textwidth}
        \centering
        \includegraphics[width=\linewidth]{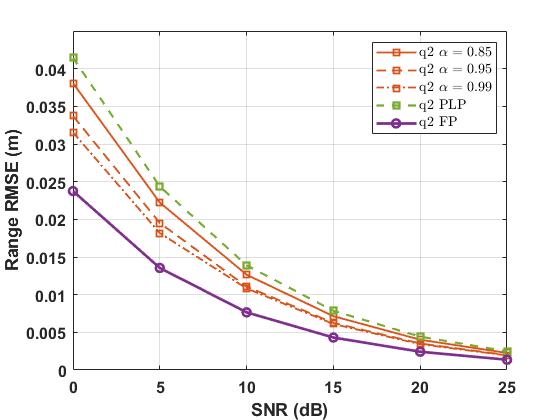}
        \caption{Range RMSE.}
        \label{q2_range}
    \end{subfigure}

    \caption{Q2 performance analysis: RMSE vs SNR for all proposed schemes, $\rho =0.3$ and $\eta_{min} = 0.85$.}
    \label{fig:q2_results}
\end{figure*}

\begin{figure*}[!t] 
    \centering
    \begin{subfigure}[b]{0.32\textwidth}
        \centering
        \includegraphics[width=\linewidth]{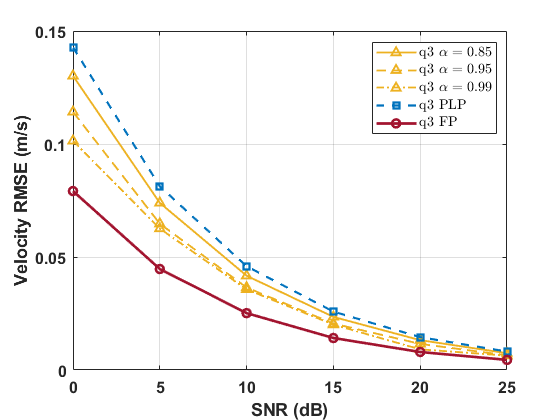}
        \caption{Velocity RMSE.}
        \label{q3_v}
    \end{subfigure}
    \hfill
    \begin{subfigure}[b]{0.32\textwidth}
        \centering
        \includegraphics[width=\linewidth]{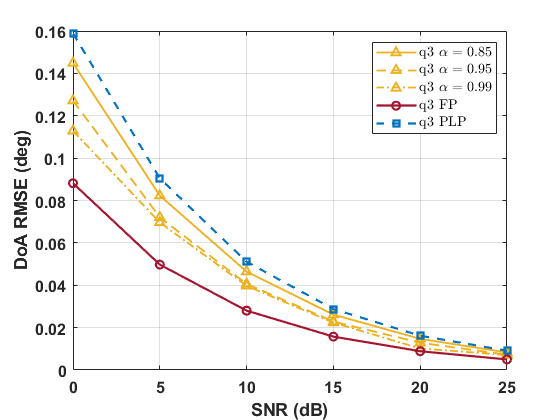}
        \caption{DoA RMSE.}
        \label{q3_doa}
    \end{subfigure}
    \hfill
    \begin{subfigure}[b]{0.32\textwidth}
        \centering
        \includegraphics[width=\linewidth]{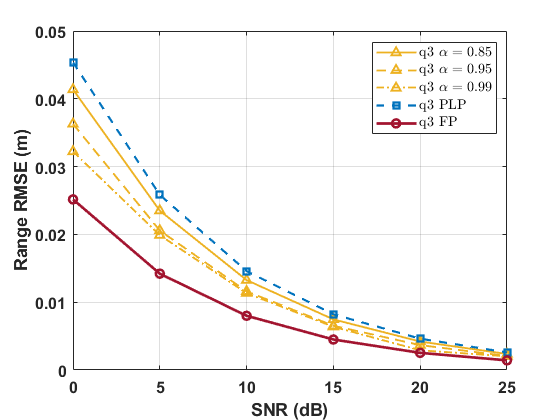} 
        \caption{Range RMSE.}
        \label{q3_range}
    \end{subfigure}

    \caption{Q3 performance analysis: RMSE vs SNR for all proposed schemes, $\rho =0.3$ and $\eta_{min} = 0.85$.}
    \label{fig:q3_results}
\end{figure*}

\begin{figure*}[!t] 
    \centering
    \begin{subfigure}[b]{0.32\textwidth}
        \centering
        \includegraphics[width=\linewidth]{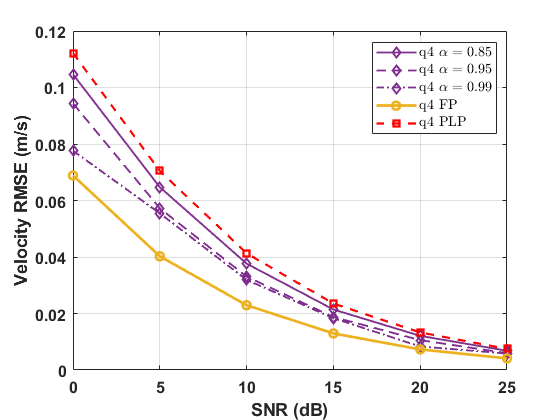}
        \caption{Velocity RMSE.}
        \label{alpha_v}
    \end{subfigure}
    \hfill
    \begin{subfigure}[b]{0.32\textwidth}
        \centering
        \includegraphics[width=\linewidth]{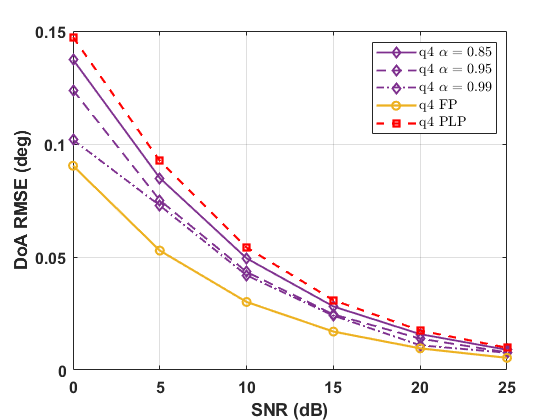}
        \caption{DoA RMSE.}
        \label{alpha_DoA}
    \end{subfigure}
    \hfill
    \begin{subfigure}[b]{0.32\textwidth}
        \centering
        \includegraphics[width=\linewidth]{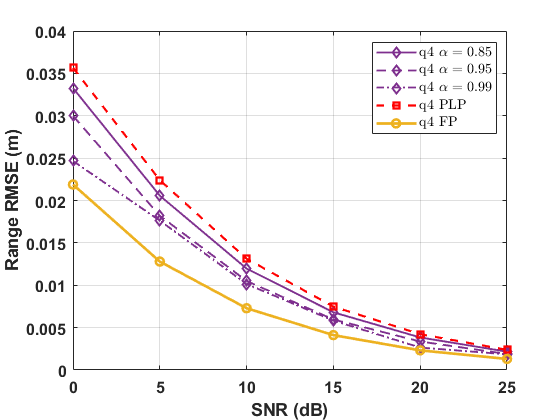}
        \caption{Range RMSE.}
        \label{range_q4}
    \end{subfigure}

    \caption{Q4 performance analysis: RMSE vs SNR for all proposed schemes, $\rho =0.3$ and $\eta_{min} = 0.85$.}
    \label{fig:q4_results}
\end{figure*}

We conduct a similar sensitivity study by sweeping $\alpha$ and reporting the corresponding  RMSE (DoA, range, and velocity) and normalized transmit power, as shown in Figs. \ref{alpha_pt}, \ref{fig:q1_results}, \ref{fig:q2_results}, \ref{fig:q3_results}, and \ref{fig:q4_results} . The results prove that the proposed scheme follows a predictable manner since increasing $\alpha$ yields reductions in both power transmission and RMSE due to improved gating efficiency. The $\alpha =0.95$ adopted in the manuscript is selected from a broad region where performance trends are stable, indicating that the proposed framework does not rely on fine parameter tuning. To keep a clear representation of the curves within the figures, we provide the results for each target on a separate figure.


\begin{figure}[!t] 
    \centering
    \begin{subfigure}[b]{\columnwidth}
        \centering
        \includegraphics[width=0.75\linewidth]{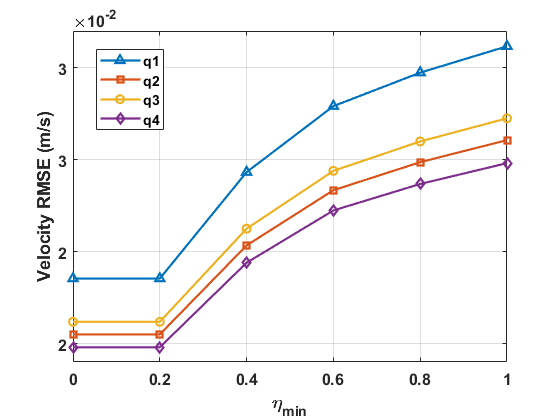}
        \caption{Velocity RMSE vs $\eta_{min}$.}
        \label{velocity}
    \end{subfigure}
    
    \vspace{10pt} 
    
    \begin{subfigure}[b]{\columnwidth}
        \centering
        \includegraphics[width=0.75\linewidth]{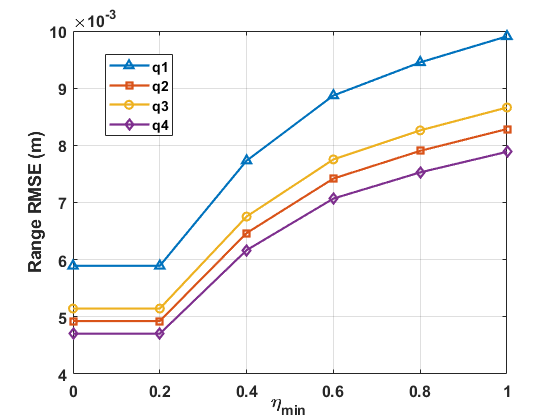}
        \caption{Range RMSE vs $\eta_{min}$.}
        \label{range}
    \end{subfigure}
    
    \vspace{10pt}
    
    \begin{subfigure}[b]{\columnwidth}
        \centering
        \includegraphics[width=0.75\linewidth]{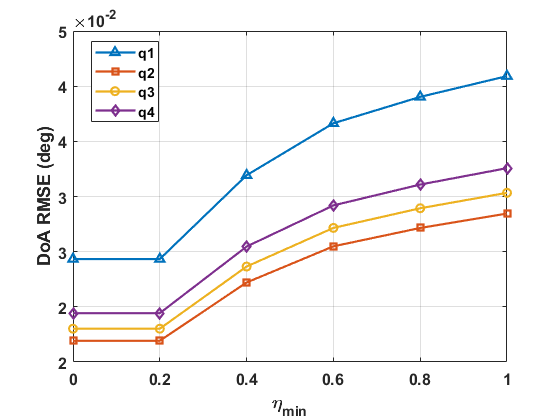}
        \caption{DoA RMSE vs $\eta_{min}$.}
        \label{DoA}
    \end{subfigure}

    \caption{Performance analysis vs $\eta_{min}$ at SNR = 15 dB, $\rho =0.3$ and $\alpha = 0.95$.}
    \label{fig:eta_results}
\end{figure}

\subsubsection{The minimum posterior confidence $\eta_{\min}$}

The parameter $\eta_{\min}$ specifies the minimum posterior confidence required to skip an active sensing update. A smaller $\eta_{\min}$ enforces more conservative sensing behavior, while larger values enable more aggressive sensing skipping and energy savings.

\begin{figure}[!t]
    \centering
\includegraphics[width=0.85\linewidth]{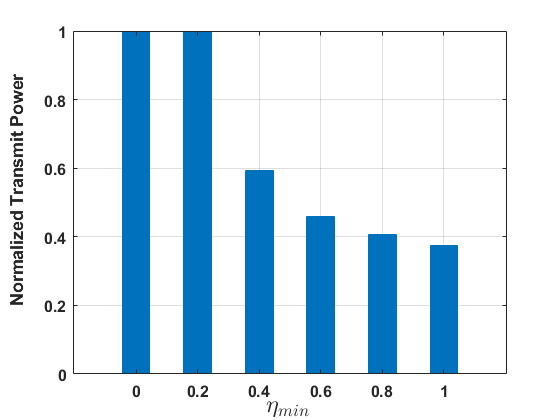}
    \caption{Normalized transmit power vs $\eta_{min}$ at SNR = 15dB, $\rho =0.3$ and $\alpha = 0.95$.}
    \label{Pt}
\end{figure}

To examine its impact, we sweep $\eta_{\min}$ over a wide range and evaluate the resulting RMSE (DoA, range, and velocity) together with the normalized transmit power under a fixed SNR. The corresponding results are provided in Figs. \ref{Pt}, \ref{velocity}, \ref{range}, and \ref{DoA}. The results show a clear trade-off: increasing $\eta_{\min}$ initially reduces normalized transmit power, whereas overly large values eventually degrade estimation performance due to insufficient measurement updates.

Finally, we emphasize that the performance gains of the proposed ASP framework are not driven by a particular choice of $\eta_{\min}$ or $\alpha$, but rather come from the underlying confidence-aware sensing mechanism. In particular, the energy efficient performance of ASP consistently outperforms the PLP and FP schemes across all values of $\alpha$ while for all values of $\eta$ compared with FP and for $\eta >0.4$ compared to PLP, as demonstrated in Fig. \ref{alpha_pt} and Fig. \ref{Pt}, respectively. 

In terms of sensing performance, the ASP  exhibits a slight degradation compared to the FP baseline. The reason for this is that ASP skips some of the epochs, which means that sensing capability is reduced. However, this degradation is kept relatively low and acceptable due to the proposed skip-aware design. Specifically, sensing deactivation is not a hard shutoff even during skipped epochs; a nonzero safety illumination floor is alwyes maintained.

However relaxed sensing holds only when the estimation confidence is sufficient and the innovation stays within the gating region; otherwise, full sensing will be activated. This approach allows sensing resources to be reduced only when reliable tracking can be maintained. Thus, achieving significant energy savings with negligible impact on tracking performance. On the other hand, the PLP baseline shows consistently higher RMSE than both FP and ASP. This is because PLP does not change the sensing activation depending on the current state of tracking and filter consistency, which means sensing updates may be skipped even when estimation uncertainty is increasing.

As for energy efficiency, Fig.~\ref{alpha_pt} shows that the FP scheme consistently needs the highest transmit power due to its continuous sensing process, while the PLP scheme achieves moderate energy savings by reducing the sensing frequency.
The ASP achieves the lowest normalized transmit power across the entire SNR range. By selectively activating sensing based on EKF confidence and NIS gating, and maintaining only a minimal safety illumination floor during skipped epochs.

\subsubsection{The safety illumination factor $\rho$}
\begin{figure*}[!t] 
    \centering
    
    \begin{subfigure}[b]{0.32\textwidth}
        \centering
        \includegraphics[width=\linewidth]{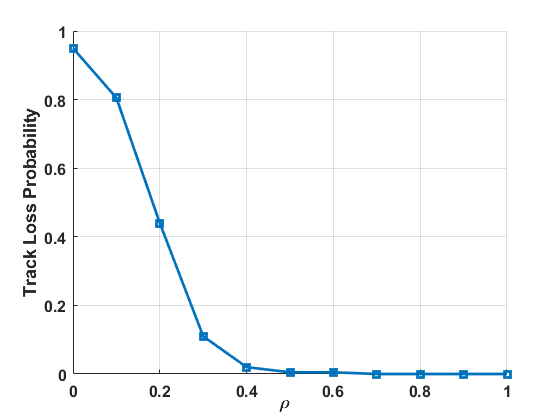}
        \caption{Tracking loss probability vs $\rho$.}
        \label{Track1}
    \end{subfigure}
    \hfill
    \begin{subfigure}[b]{0.32\textwidth}
        \centering
        \includegraphics[width=\linewidth]{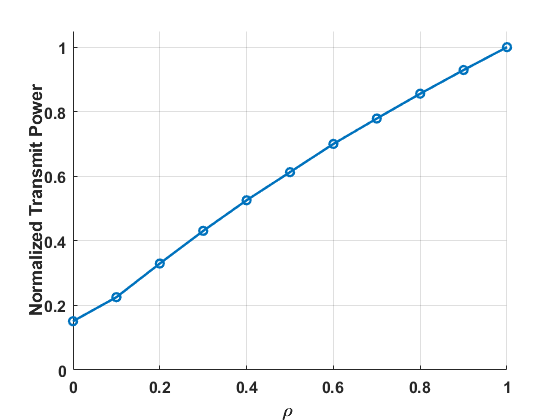}
        \caption{Normalized transmit power vs $\rho$.}
        \label{track2}
    \end{subfigure}
    \hfill
    \begin{subfigure}[b]{0.32\textwidth}
        \centering
        \includegraphics[width=\linewidth]{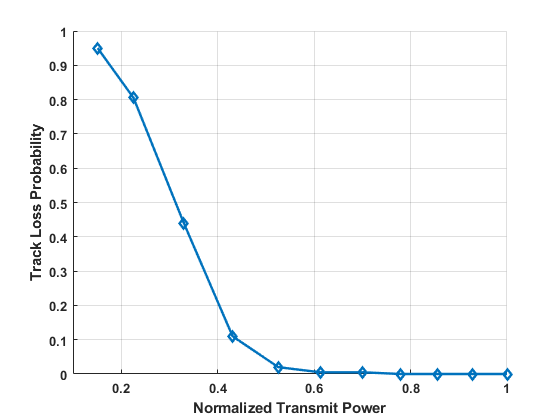}
        \caption{Tracking loss probability vs normalized transmit power.}
        \label{track3}
    \end{subfigure}

    \caption{Tracking loss performance analysis for ASP.}
    \label{fig:tracking_results}
\end{figure*}

The safety illumination factor $\rho$ is essential for achieving a balance between tracking reliability and energy efficiency within the proposed ASP framework. To quantify its impact, Fig.~\ref{fig:tracking_results} reports the track-loss probability and normalized transmit power as functions of $\rho$.

Two trends can be noted here. Firstly, the track-loss probability rapidly decreases as $\rho$ increases from small values, as shown in Fig.~\ref{Track1}. The reason for such behavior is that in the case of small $\rho$ values, the reduction of sensed illumination in the skipping epoch makes the process unreliable and leads to frequent track divergences, while with an increase in $\rho$ the tracking performance improves, but beyond a certain moderate value, the decrease in the track loss probability becomes insignificant due to its rapid saturation.  

Secondly, the normalized transmit power grows monotonically as $\rho$ increases, as illustrated in Fig.~\ref{track2}. The explanation behind this trend lies in the fact that the increased safety illumination floor forces more frequent probe transmission even during high-confidence tracking epochs. Hence, with an increase of  $\rho$, there is less energy efficiency.

The combined effect of these trends is presented in Fig.~\ref{track3}. In particular, the optimal value of $\rho$ can be found at $\rho=0.3$. Such a choice allows improving the tracking performance and maintaining energy efficiency as seen from Fig. \ref{Track1}. Increasing $\rho$ beyond this point provides only marginal gains in tracking robustness, at the cost of a significant increase in transmit power.\par

\begin{figure}[!t]
\centering
\includegraphics[width=.85\linewidth]{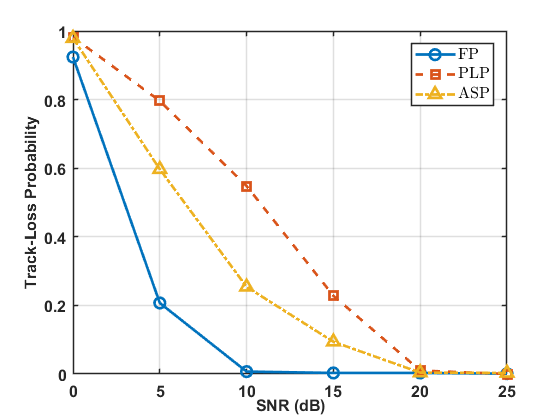}
  \caption{Tracking loss probability vs SNR (dB).}
  \label{tracking_all}
\end{figure}

In Fig.\ref{tracking_all}, the dependence of the track-loss probability on SNR for all the schemes with the ASP value $\rho = 0.3$ is given. As expected, the FP scheme achieves the lowest track-loss probability throughout the whole SNR range since continuous measurement is performed by this scheme and, hence, it achieves the best performance. However, this performance comes at the cost of the highest energy consumption. In contrast, the PLP scheme provides the highest track-loss probability. Due to a fixed periodic sensing schedule, which may lead to missed updates during critical periods where estimation uncertainty increases. As a result, tracking errors accumulate and cause frequent track divergence. The ASP scheme significantly improves tracking robustness compared to PLP, especially at low and moderate SNR values due to strategic activation-based.  Overall, it can be concluded from the findings above that the ASP achieves a balance between tracking reliability and energy efficiency by filling the gap between PLP and FP schemes.

\subsection{Communication Performance}

\begin{figure}[H]
\centering
\includegraphics[width=.85\linewidth]{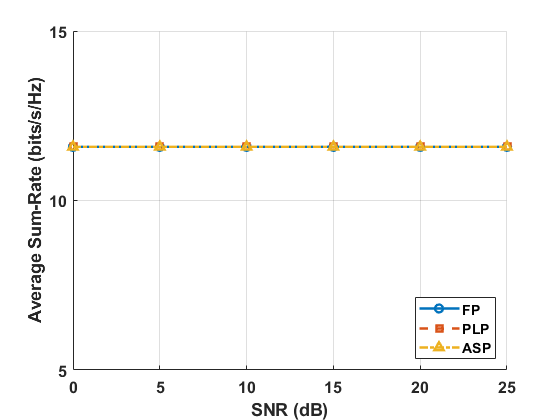}
  \caption{Average sum rate vs SNR.}
  \label{communication}
\end{figure}

Fig.~\ref{communication} shows the average sum-rate versus SNR for all considered schemes. It can be observed that the sum-rate remains nearly constant across the entire SNR range and is identical for all schemes. This behavior is expected due to the adopted minimum-power beamforming formulation, which enforces a fixed per-user SINR constraint rather than maximizing throughput. Specifically, the optimization allocates only the minimum required transmit power to satisfy the SINR targets. As a result, improvements in channel conditions (i.e., increasing SNR) are translated into reduced transmit power rather than increased data rates.
Importantly, the identical sum-rate across FP, PLP, and ASP confirms that all schemes preserve the same communication QoS. This validates that any performance differences observed in the subsequent results are not due to communication degradation, but rather stem from differences in sensing and energy management strategies.

\begin{figure}[!t]
\centering
\includegraphics[width=.85\linewidth]{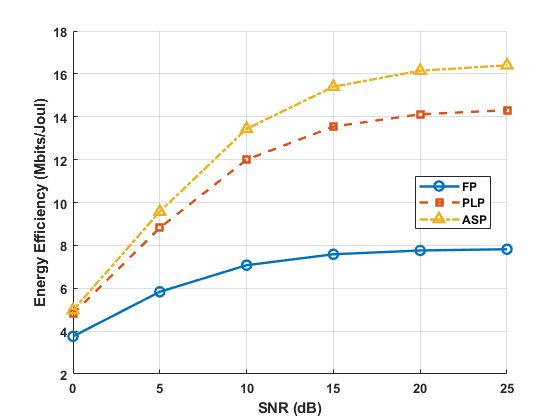}
  \caption{Energy efficiency vs SNR.}\label{EE}
\end{figure}

Having established that all schemes achieve identical communication performance. Fig.~\ref{EE} illustrates the EE performance of the considered schemes versus SNR, highlighting the inherent trade-off between communication performance and power consumption. Unlike conventional rate-maximization approaches, the proposed framework enforces a minimum SINR constraint while minimizing transmit power, such that improvements in channel conditions are reflected in reduced power consumption rather than increased data rates. As observed, the proposed ASP scheme consistently achieves the highest EE across all SNR values, followed by PLP, while FP exhibits the lowest efficiency. This trend underscores the benefit of avoiding unnecessary sensing activity. Specifically, as the SNR increases, the transmit power required to satisfy the SINR constraints decreases, resulting in a monotonic increase in EE for all schemes.

Furthermore, the performance gain of ASP becomes more pronounced at higher SNR levels. This is attributed to its ability to dynamically adapt sensing activity based on EKF confidence and NIS gating, thereby skipping redundant probing while maintaining reliable tracking. In contrast, FP enforces continuous sensing regardless of the tracking state, leading to excessive power consumption. Although PLP reduces the probing frequency, its fixed periodic strategy lacks adaptability, resulting in suboptimal energy utilization. 

 In summary, the findings clearly indicate that the ASP greatly improves the energy efficiency without sacrificing communication performance, making it suitable for green ISAC.

\section{Conclusion}
A skip-aware ISAC sensing technique based on the combination of an EKF posterior confidence measure and NIS test for sensing control was proposed in this paper. As noted, full-power probing is not always necessary for maintaining high tracking performance. For that reason, the proposed approach developed here provides the ability to loosen sensing requirements while retaining minimal illumination in order not to break tracking. Thus, the resulting framework explicitly accounts for the trade-off between sensing reliability and energy consumption by adapting the probing activity to the instantaneous tracking status. Simulation results demonstrate that the method significantly reduces transmit power relative to baselines, preserves communication performance, and acceptable degradation of tracking accuracy. Moreover, the proposed approach achieves a favorable energy-tracking trade-off, highlighting the benefit of confidence-aware sensing decisions. These results confirm that reliable tracking can be maintained without continuous sensing, provided that observability is preserved through a minimum illumination floor. From a system perspective, the proposed strategy enables a shift from always-on sensing to adaptive, information-driven operation. Future work will extend the framework to extended target tracking and investigate antenna selection co-designed with the skip-aware policy to enhance energy efficiency further.

\bibliographystyle{IEEEtran}
\bibliography{references}

\end{document}